\documentclass{nature}
\usepackage{lineno}
\usepackage{float}
\usepackage{graphicx,color}
\usepackage{dcolumn}
\usepackage{amsmath}
\usepackage{epsfig}
\title{Unusual evolution of $B_{c2}$ and $T_c$ with inclined fields in restacked TaS$_2$ nanosheets}

\author{Yonghui Ma$^{1,3,4}$, Jie Pan$^{2,4}$, Chenguang Guo$^{2,4}$, Xuan Zhang$^{1}$, Lingling Wang$^{1}$, Tao Hu$^{1,3}$, Gang Mu$^{1,3,*}$, Fuqiang Huang$^{2,3,\dagger}$, and Xiaoming Xie$^{1,3,4}$}

\begin{document}
%\linenumbers
\maketitle

\begin{abstract}
Recently we reported an enhanced superconductivity in restacked
monolayer TaS$_2$ nanosheets compared with the bulk TaS$_2$,
pointing to the exotic physical properties of low dimensional
systems. Here we tune the superconducting properties of this system
with magnetic field along different directions, where a strong Pauli
paramagnetic spin-splitting effect is found in this system.
Importantly, an unusual enhancement as high as 3.8 times of the upper critical field
$B_{c2}$, as compered with the Ginzburg-Landau (GL)
model and Tinkham model, is observed under the inclined external magnetic field.
Moreover, with the out-of-plane field fixed, we find that the
superconducting transition temperature $T_c$ can be enhanced by
increasing the in-plane field and forms a dome-shaped phase
diagram. An extended GL
model considering the special microstructure with wrinkles was proposed to describe the results. The restacked crystal structure without
inversion center along with the strong spin-orbit coupling may also play
an important role for our observations.
\end{abstract}

\footnotetext[1]{State Key Laboratory of Functional Materials for
Informatics, Shanghai Institute of Microsystem and Information
Technology, Chinese Academy of Sciences, Shanghai 200050, China.
$^2$State Key Laboratory of High Performance Ceramics and Superfine
Microstructure, Shanghai Institute of Ceramics, Chinese Academy of
Sciences, Shanghai, 200050, China. $^3$Center for Excellence in Superconducting Electronics (CENSE), Chinese Academy of Sciences, Shanghai 200050, China.
$^4$University of Chinese Academy of Sciences, Beijing 100049,
China. Correspondence and requests for materials should be addressed
to G.M. (email: mugang@mail.sim.ac.cn) and F.Q.H. (email:
huangfq@mail.sic.ac.cn). Y.H.M. and J.P. contributed equally to this
work.}

\clearpage

\section*{Introduction}
Superconductivity in low-dimensional systems was investigated
extensively recently, due to the fertile physical phenomenon and
exotic properties~\cite{Reyren1196,Gozar,FeSe,Saito}. At
present, the gate of this research field has just been opened and
more interesting phenomena are waiting to be explored. Because of the strong spin-orbit
coupling, superconducting transition metal dichalcogenides (TMDs)
are investigated intensively in the two-dimensional (2D) limit in recent
years~\cite{NbSe2-1,MoS2-1,MoS2-2,NbSe2-2}. A clear
enhancement of the in-plane upper critical field was frequently
reported in these materials, which was interpreted by the Zeeman-protected Ising
superconductivity mechanism. Using a chemical exfoliation method, we
have obtained the monolayer TaS$_2$ nanosheets, which were assembled
layer-by-layer by vacuum filtration~\cite{TaS2,C7TC00838D}. Such a
restacked material shows superconductivity with $T_c$ ($\sim$ 3.2 K)
several times higher than the pristine bulk 2H-TaS$_2$, which
supplies a significant platform for studying the intrinsic physical
properties of unconventional superconductivity in TMDs. Such an enhancement of $T_c$
is consistent with a previous work by E. N. Moratalla et al~\cite{TaS2-1}, although they didn't
reach the one-layer limit, and finally confirmed by other two groups by mechanically exfoliating TaS$_2$ to monolayer~\cite{Yang,TaS2-2}.
The enhancement of $T_c$ was believed to originate from the suppression of the charge-density wave and the increase of the density of
states by the process of thickness reduction.~\cite{TaS2,Yang}
However, an in-depth investigation on the physical behaviors of the restacked TaS$_2$ is
lacking and more experiments are required at present. Magnetic
field is one of the fundamental tuning parameters to affect the
behaviors of a superconductor. In the type-II superconductors, the
magnetic field can penetrate into the bulk in the form of quantized
vortex lines when it exceeds the lower critical field
$B_{c1}$~\cite{Abrikosov}. In addition, various pair-breaking effect, including the orbital
and Zeeman type, can be induced by the magnetic field.

Here we present a detailed investigation on the Abrikosov vortex
phase of the above-mentioned superconductor, restacked TaS$_2$
nanosheets, by measuring the conducting properties with magnetic
fields along different directions. The in-plane upper critical field $B_{c2}^{ab}$ is clear larger than the Pauli
paramagnetic limiting fields $B_P$, indicating a strong Pauli
paramagnetic spin-splitting effects in this material. Importantly,
the angle dependence of the upper critical field deviates severely
from the Ginzburg-Landau (GL) model and Tinkham model. Moreover, the value of $T_c$ is
found to increase with the in-plane field $B_{\parallel ab}$ under
a fixed out-of-plane field $B_{\parallel c}$. Both the intrinsic and extrinsic origins for the observations were discussed and we found that the highly
noncentrosymmetric crystal structure, the strong spin-orbit
coupling, and the special microstructure with wrinkles are important factors for the unusual behaviors we observed.

%the superconducting transition temperature $T_c$ can be suppressed
%by the field induced pair-breaking effect, with either the orbital
%or the Zeeman type. In anisotropic materials, the efficiency of
%$T_c$ suppression can be rather different with field along different
%directions: typically $T_c$ is suppressed more quickly with $B//c$
%than that with $B//ab$, if the electrons conduct within the
%$ab$-palne.

\section*{Results}

Details regarding the samples preparation and resistance
measurements are given in the Methods section. By a careful
characterization using combining methods, the structure of the
restacked TaS$_2$ was determined and reported in our previous paper~\cite{TaS2}. The inter-layer spacing is close
to bulk 2H-TaS$_2$, while the 2H symmetry has been broken after the
restacking process because of rotations between different layers
(see Fig. S1). In such a structure, both the in-plane inversion
symmetry in each individual layer and the global inversion symmetry
are broken. As a consequent, the inversion symmetry breaking will be
severer than the bulk, monolayer, and few layered TaS$_2$
materials. The morphology of the samples were checked by the scanning electron microscope (SEM)
and scanning transmission electron microscope (STEM) measurements. Wrinkles can be seen on the flat surface and oxides exist only at the edge of the sample with the
width of several nanometers. The content of lithium ions was found to be below the minimum detection limit by the inductively coupled plasma optical emission spectrometer (ICP-OES) measurement.
A clear Meissner effect can be seen from the magnetic susceptibility measurements (see Fig. S3d).
These characterizations can be found in the Supplementary information.

The resistive transitions of one sample (denoted as \#1) measured in
magnetic fields for both orientations $B\parallel ab$ and $B\parallel c$ are
shown in Figs. 1a and b. Clear different efficiencies for the
suppression of superconductivity, revealing the anisotropy of the present material, can be seen by comparing the two
figures. To determine the upper critical fields $B_{c2}$, a
criterion of $90\%$ of the normal state resistivity ($\rho_n$) is used
and the results are shown in the Fig. 1c for the two orientations.
Instead of the square root behavior for the in-plane upper critical
field ($B_{c2}^{ab} \sim \sqrt{1-T/T_c}$) expected for the 2D
superconductors~\cite{NbSe2-1,MoS2-1,MoS2-2,NbSe2-2,Yang}, an opposite tendency with a positive curvature is
observed. This has been found to be a universal feature of anisotropic three-dimensional (3D) layered
superconductors~\cite{PhysRevB.13.3843,CaFeAsF}. This reflects the
influence of inter-layer coupling on the in-plane upper critical
field of our samples, although such an inter-layer stacking manner
doesn't affect $T_c$. The value of $B_{c2}$ at zero temperature can
be estimated using the Werthamer-Helfand-Hohenberg
relation~\cite{WHH} $B_{c2} =-0.693\times dB_{c2}(T)
/dT|_{T_c}\times T_c$ after the slope $dB_{c2}(T) /dT|_{T_c}$ is
obtained from Fig. 1c. In addition, the paramagnetic limiting
field $B_P$ has a simple relation with $T_c$, $B_P = 1.84\times T_c$
based on the conventional BCS theory~\cite{ParaLimit}. The resultant
values for the three characteristic fields $B_{c2}^{ab}$ (in-plane
$B_{c2}$), $B_{c2}^{c}$ (out-of-plane $B_{c2}$) and $B_P$ are denoted
by arrows in Figure 1c and summarized in Table 1, from which the
anisotropy of upper critical field $\Gamma = B_{c2}^{ab}/B_{c2}^{c}
= 11$ is obtained. This value is larger than most of the iron-based
superconductors and the copper-based superconductor
YBCO~\cite{CaFeAsF,Gamma1}. Moreover, a clear relative relation
$B_{c2}^{c}<B_P<B_{c2}^{ab}$ can be deduced.

Field-angle resolved experiments were performed by measuring the
field ($B$) and angle ($\theta$) dependence of resistivity at a fixed
temperature 2.2 K. As shown in Fig. 2a, one can see how the
resistivity is triggered by the field from zero to finite values and
saturates gradually at high fields. In order to determine the
precise onset superconducting transition point, which reflects the
information of upper critical fields $B_{c2}$, we show the first
derivative $d\rho/dB$ of four typical curves in Fig. 2b. As indicated
by the arrows, the onset transition point is defined by the
characteristic field where the value of $d\rho/dB$ begins to increase
clearly. The characteristic points determined in Fig. 2b are
represented by arrows in Fig. 2a, the connection of which forms a
slightly inclined straight line as shown by the blue dashed line.
Based on this line reflecting the normal states resistance $\rho_n$, a
criterion of 90\%$\rho_n$, as revealed by the black dashed line, is
adopted to define the upper critical fields. The crossing points
between this black dashed line and the data curves in Fig. 2a
determine the upper critical field at different angles
$B_{c2}(\theta)$. Angle dependence of $B_{c2}(\theta)$ normalized
by $B_{c2}^c$ is shown in Fig. 2c. One can see the detailed
evolution of $B_{c2}(\theta)/B_{c2}^c$ versus $\theta$. In order to
quantitatively evaluate such an angle dependent variation, we employ two theoretical models, the 3D GL model and 2D
Tinkham model~\cite{GL,Tinkham} (see SI),
and plot the curves based on them for comparison. These two models show
slight differences near $\theta=0^o$, which is usually used to
distinguish the 2D superconductivity in monolayer or interfacial
systems~\cite{MoS2-1,MoS2-2,2D1}. However, the difference between our experimental data and the two
models is much larger, showing a great enhancement of upper critical
field in a wide angle range. In the strong
anisotropic system, the perpendicular component of the the upper
critical field $B_{c2}(\theta)sin\theta$ is expected to be dominant
in the low angle range since the in-plane magnetic field is not important in terms of suppressing $T_c$. So we show this component normalized by
$B_{c2}^c$ in Fig. 2d, where a broad peak-shaped experimental
curve with the maximum enhancement of 3.8 times is observed.

Actually such an upper-critical-field-enhancement effect can induce
very fascinating features in $T_c$ in the mixed state. Here we adopt
a different mode of measurements: $\rho-T$ curves are measured with the
out-of-plane field $B_{\parallel c}$ fixed and the in-plane field
$B_{\parallel ab}$ increasing, as schematized in the inset of Fig.
3b. The current is applied in the direction perpendicular to $B_{\parallel ab}$. This method has been used to identify the 2D or interfacial
superconductivity, where the curves will overlap with each other
because $B_{\parallel ab}$ is not important in an extremely 2D
superconductor~\cite{2D1,2D2}. While in an anisotropic system with 3D
features, $T_c$ will be reasonably suppressed by $B_{\parallel ab}$
(see section 7 of SI). In Fig. 3a we show a typical
set of data with $B_{\parallel c} = 0.4$ T on another sample denoted
as \#2 with a similar $T_c$ as \#1. The behavior is uniquely
different from the above-mentioned two categories. The
superconducting transition is enhanced clearly following by a
suppression with the increasing of $B_{\parallel ab}$. Three
criterions, 10\%$\rho_n$, 50\%$\rho_n$, and 90\%$\rho_n$, are employed to determine the critical transition
temperature $T_c$ and the
results are shown in Fig. 3b. All the three curves show the
dome-like features, confirming that it is an intrinsic property
rather than a magnetic flux-related behavior. The dome-like behavior
may reveal the presence of competing between some unconventional
effect and the pairing-breaking effect induced by magnetic field.
Figure 3c summarizes the effect of
$B_{\parallel ab}$ on $T_c$ at different $B_{\parallel c}$, where
$T_c$ is determined using the criterion 50\%$\rho_n$. The maximum $T_c$ for the three curves emerges at the same ratio $B_{\parallel ab}/B_{\parallel c}$,
indicating a characteristic angle $\theta \sim 20 ^o$, along which the detrimental effect of magnetic field is
mostly suppressed.

\section*{Discussion}

The detailed investigations on the thickness
dependence of superconducting behaviors of 2H-TaS$_2$~\cite{Yang} supply
a good coordinate to make a comparison with our results. It is found that the critical transition temperature $T_c$
increases with the decrease of thickness and reaches 3.4 K for
monolayer TaS$_2$. This value is very close to our sample,
confirming the monolayer-features of our sample and suggesting that
the restacking process imposed on the monolayer TaS$_2$ doesn't
affect $T_c$ of this system. Moreover, the normal state
resistivity displays a $T^{2.45}$ behavior in low temperature in our
sample (see Fig. S4), corresponding to the situation between
3-layer ($\sim T^2$) and 7-layer ($\sim T^3$) for the ordered
stacked TaS$_2$~\cite{Yang}. This implies that the inter-layer
coupling in our samples shows a certain degree of influence on the
electrical transport behavior.  A more
careful examination shows that the out-of-plane upper critical field
$B_{c2}^c$ is similar to the bilayer TaS$_2$, while the in-plane
$B_{c2}^{ab}$ and the anisotropy are only one third of the bilayer
TaS$_2$. Nevertheless, $B_{c2}^{ab}$ of our samples is clearly
larger than that of the bulk samples since the latter doesn't exceed
the Pauli limit. All in all, the present samples are different from
both the monolayered and the bulk TaS$_2$. The restacked TaS$_2$ monolayers
maintain the enhanced $T_c$ (compared with the bulk material), while
lose the 2D characters and show an anisotropic 3D features. This is the
basis of the following discussions.

Field-induced superconductivity has been theoretically proposed for ferromagnetic materials and is known as Jaccarino-Peter compensation effect.~\cite{Jaccarino-Peter} In this effect, the internal
magnetic field created by the magnetic moments through the exchange interaction can be compensated by the external magnetic field and superconductivity will occur. This mechanism has been
realized in Eu-Sn molybdenum chalcogenides experimentally.~\cite{Eu-Sn} Moreover, another theory proposed by Kharitonov and
Feigelman considered the polarization of magnetic impurity spins induced by the in-plane field and predicted an enhancement of superconductivity, especially in disordered films.~\cite{Kharitonov}
Evaluating the performances of our samples, we found that neither of the two theories is applicable. First of all, our magnetization measurement shows a paramagnetic behavior in our samples (see Fig. S3(c)),
which excludes the presence of long-range magnetic moments and magnetic impurities. Secondly, the enhancement of $T_c$ by in-plane field in our samples can only be observed in the presence of out-of-plane field,
which is rather different from the case of the above-mentioned scenarios.

Intuitively, the theoretical proposal~\cite{Lebed1,Lebed2,Lebed3} predicting a field-induced
triplet component in the order parameter of the singlet
superconductors due to the Pauli paramagnetic spin-splitting
effect is very consistent with our
observation shown in Fig. 2(d). According to their arguments, such
an enhancement of $B_{c2}(\theta)$ should be conspicuous in
anisotropic superconductors with $B_{c2}^{c}\ll B_P\ll B_{c2}^{ab}$.
However, we note that such an exotic behavior is absent in so many
low-dimensional TMDs with even stronger Pauli paramagnetic
spin-splitting effect~\cite{MoS2-1,MoS2-2}, which weakens the persuasion of this
interpretation. Other important factors should be considered to interpret our experiments. One important clue for exploring the physical
origination is that such a rare behavior was also observed in
restacked 1T'-MoS$_2$ (see Fig. S7) prepared with the similar
process~\cite{MoS2-3} to that used in restacked TaS$_2$. This implies that the
unique and common features of such restacked monolayered materials
are key factors for our observations. One most conspicuous
feature for such restacked monolayer materials is the
noncentrosymmetric crystal structure as mentioned in the beginning
of the Results section. It has been discussed a lot both
theoretically and experimentally that the noncentrosymmetric crystal
structure along with strong spin-orbit coupling (SOC), which also
exists in the present compound with 5d metal, is very favorable to
incur the spin-triplet component in the superconducting order
parameter~\cite{Theory1,Theory2,Theory3,Theory4,LiPtB,LiPtB2,CePtSi}. This may be one possible
origin of our observations. One positive evidence for this scenario
is that the restacked 1T'-MoS$_2$, which have a weaker SOC because
of the lighter 4d Mo element, shows an inconspicuous enhancement of
$B_{c2}(\theta)$ compared with restacked TaS$_2$ (see Fig. S7).

One possible extrinsic origination to explain our results comes from
the possible orientation mismatch or wrinkles (see Fig. S2) of the monolayer
TaS$_2$ sheets during the restacking process, which affects the
$c$-axis orientation of the restacked samples. In section 9 of SI, we carefully analyzed the influences
of different orientations on the estimation of $B_{c2}$ and found that $B_{c2}$ is mainly determined by the high-$T_c$ portion where the included angle
between the surface and field is the smallest. The presence of the low-$T_c$ portion with a larger included angle can lower the onset transition temperature slightly (see Fig. S8).
The simple GL model~\cite{GL} is not applicable any more in this case.
Considering the misaligned angle $\Delta \theta$ resulting from the wrinkles as schematically displayed in the inset of Fig. 4, we proposed an extended GL (EGL) model:

\begin{equation}\begin{split}
B_{c2}(\theta)& =[1-A(\theta)]\times \frac{B_{c2}^{c}}{\sqrt {{{\cos }^2}\Theta/{\gamma
^2}+{{\sin }^2}\Theta}},\label{eq:1}
\end{split}\end{equation}
where $\Theta = \min |\theta-\Delta\theta|$ is the smallest included angle
between the surface and field, and $A(\theta)$ is an adjusting parameter taking into account the effect of the low-$T_c$ portion with other orientations. Here we simply assume that $A(\theta)$
is proportional to the GL formula, because $A(\theta)$ will be larger when $T_c$ is more sensitive with the orientation, which can be reflected by the GL formula. Typically the misaligned angle $\Delta\theta$
has a distribution range with a maximum $\Delta\theta_{Max}$. Then we have

\begin{equation}
\Theta = \left\{\begin{aligned}
\theta-\Delta\theta_{Max},\quad\quad \theta > \Delta\theta_{Max}\\
0.\quad\quad\quad\quad\quad \theta \leq \Delta\theta_{Max}
\end{aligned}\right.
\end{equation}
By tuning the value of parameters, it is found that when $\Delta\theta_{Max} = 27^o$, this EGL model can well describe the experimental data in the range $\theta > \Delta\theta_{Max}$, as shown in Fig. 4a.
As for the case of $\theta \leq \Delta\theta_{Max}$, $B_{c2}(\theta)$ is only determined by the adjusting parameter $A(\theta)$. However, the influence factors on $A(\theta)$ are rather complicated in this
region, so we could not get a good description at present. At this stage, we could not exclude the possibility that the wrinkles have an important influence on our observations.
Even in this scenario, our results supplied an interesting prototypical system showing an unambiguous relationship between the microstructure and the physical (superconducting) performance. Moreover, this is also
valuable in designing devices for applications.

It is notable that the enhancement of $T_c$ by $B_{\parallel ab}$ at the present of $B_{\parallel c}$ is simply a natural consequence of the unusual behavior in $B_{c2} (\theta)$.
As shown in Fig. 4b and 4c, supposing a linear suppression of $B_{c2}$ with the temperature approaching $T_c$: $B_{c2}(T) \sim 1-T/T_c$, angle dependence of $B_{c2}(\theta)$ at other temperatures can be
derived from the data at 2.2 K for both the experimental data and GL model, respectively. The total values of field in the experiment of Fig. 3, $\sqrt{B_{\parallel ab}^2+B_{\parallel c}^2}$, are also plotted
for comparison. The intersections between the two sets of curves indicate the values of $T_c$. With the decrease of $\theta$, the intersection in Fig. 4b moves to high temperatures until the orientation
around $20-30^o$, and then moves to low temperatures for both cases with $B_{\parallel c}=0.2$ T and $B_{\parallel c}=0.4$ T. This reproduces the variation of $T_c$ with angle as shown in Figs. 3b and 3c.
In the normal situation as shown in Fig. 4c, the value of $\sqrt{B_{\parallel ab}^2+B_{\parallel c}^2}$ evolves along the tracing of GL model of a certain temperature in a rather wide angle range and moves
to low temperature slightly in low angle region.

To summarize, we have measured angle-resolved electrical
resistivity of the restacked TaS$_2$ nanosheets under
magnetic field along different directions. It is found that Pauli paramagnetic limit is broken through in the mixed state, placing the present superconducting system in the environment suffering a
strong Pauli paramagnetic spin-splitting effect. A clear enhancement
of the upper critical field, as compared with the GL model and Tinkham model,
is observed in inclined magnetic field. The critical transition temperature can be enhanced by the in-plane field at the presence of the out-of-plane field. Our analysis indicates
that the highly noncentrosymmetric crystal structure, the strong
spin-orbit coupling and the special microstructure with wrinkles are important factors for the mechanism
of the unusual behaviors we observed.

\begin{methods}

\subsection{Sample preparation.}

The restacked TaS$_2$ nanosheets were obtained by a chemical
exfoliation method followed by the vacuum
filtration~\cite{TaS2,C7TC00838D}. Firstly 2H-TaS$_2$ powders were
prepared by the solid-state reaction. Then the Li$_x$TaS$_2$
powders were synthesized by soaking as-prepared 2H-TaS$_2$ powders
in n-butyl lithium solution. The as-prepared Li$_x$TaS$_2$ crystals
were exfoliated in distilled water. The redox reaction occurs at
this stage. The obtained colloidal solution is composed of TaS$_2$
monolayers and is rather stable. The restacked TaS$_2$ nanosheets
were obtained from the vacuum filtration of the colloidal
suspension. The restacked TaS$_2$ samples are stable in air within 24 hours (see Fig. S2).

\subsection{Morphology characterization.}
The morphology of restacked TaS$_2$ nanosheets was characterized by JSM-6510 scanning electron microscope (SEM).
The atomic structures were observed through JEOL ARM-200F High-angle annular dark field (HAADF) scanning transmission
electron microscope (STEM).

\subsection{Resistance measurements.}
The restacked samples for the electrical transport measurements have the thickness of about 1 $\mu$m.
The electrical transport data were collected by the standard
four-probe method with magnetic field rotating in the plane
perpendicular to the electric current. $\theta$ denoted the included
angle between external field $B$ and the $ab$-plane of the crystal.
The applied electric current is 10 $\mu$A when carrying out the resistivity
measurements and up to 50 mA for the I-V measurements.

\subsection{Data availability.}
All relevant data are available from the corresponding author.
\end{methods}

%% Put the bibliography here, most people will use BiBTeX in
%% which case the environment below should be replaced with
%% the \bibliography{} command.

% \begin{thebibliography}{1}
% \bibitem{dummy} Articles are restricted to 50 references, Letters
% to 30.
% \bibitem{dummyb} No compound references -- only one source per
% reference.
% \end{thebibliography}

%% Here is the endmatter stuff: Supplementary Info, etc.
%% Use \item's to separate, default label is "Acknowledgements"
\section*{Acknowledgments}
This work is supported by the Youth Innovation Promotion Association of the
Chinese Academy of Sciences (No. 2015187), the National Natural Science Foundation of
China (No. 11204338), and the ``Strategic Priority Research Program (B)"
of the Chinese Academy of Sciences (No. XDB04040300).

\section*{Author contributions}
G.M. and F.Q.H. designed the experiments. Y.H.M. performed the
measurements. J.P. and C.G.G. synthesized the samples. G.M. analysed
the data. Y.H.M., J.P., C.G.G. X.Z., L.L.W., T.H., G.M., F.Q.H., and
X.M.X. discussed the results. G.M. and F.Q.H. wrote the paper.
X.M.X. supervised the work.

\section*{Additional information}
\subsection{Supplementary information}
accompanies the paper on the npj Quantum Materials website.

\subsection{Competing Interests:}
The authors declare no competing financial and non-financial interests.

\clearpage

\section*{Figure captions}

\begin{figure*}
\includegraphics[width=0.95\textwidth,bb=3 4 1298 542]{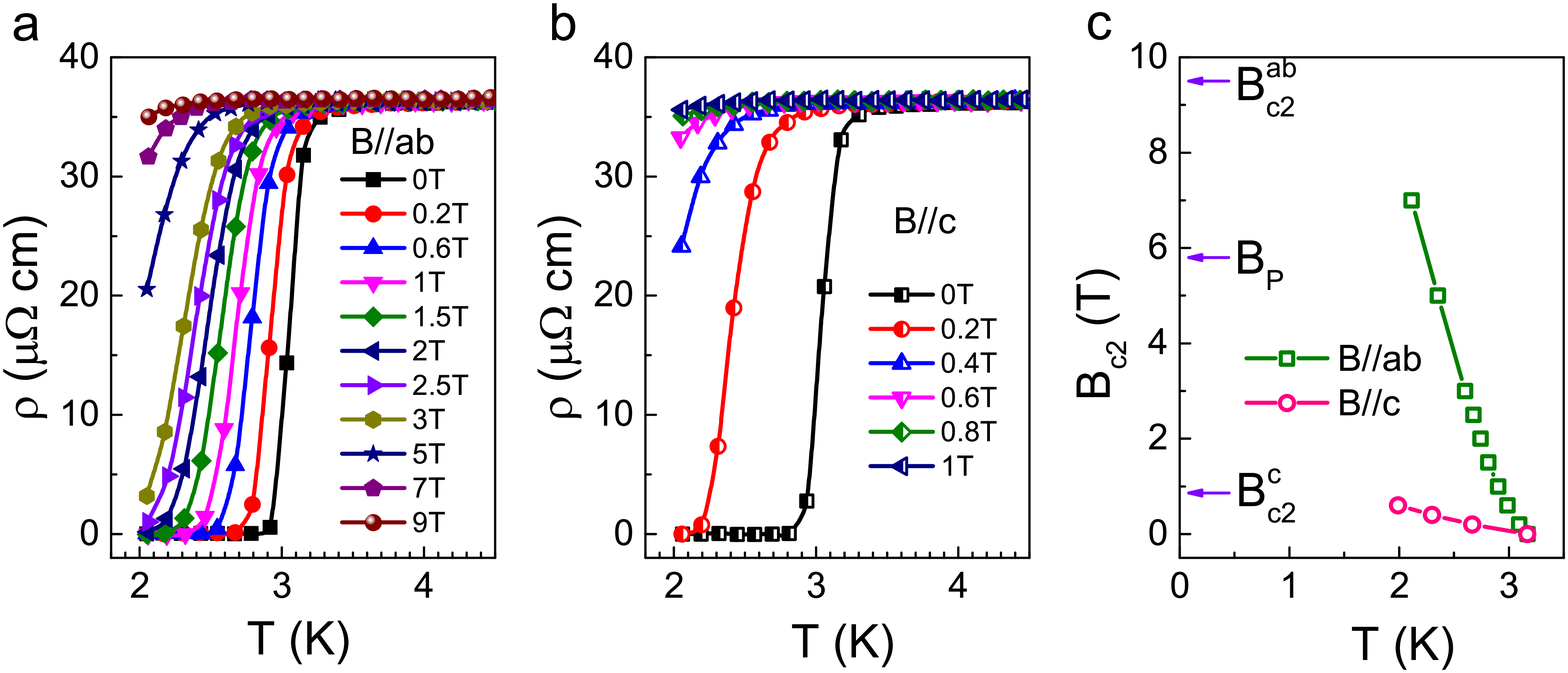}
\caption {Temperature dependence of the resistive transitions under
magnetic field for the sample \#1. (a) $B\parallel ab$. (b)
$B\parallel c$. (c) $B_{c2}-T$ phase diagram obtained using the
criterion 90\%$R_n$. The three characteristic fields $B_{c2}^{ab}$,
$B_{c2}^{c}$ and $B_{P}$ are indicated by the arrows in this
figure.} \label{fig1}
\end{figure*}

\begin{figure*}
\includegraphics[width=0.95\textwidth,bb=3 4 821 650]{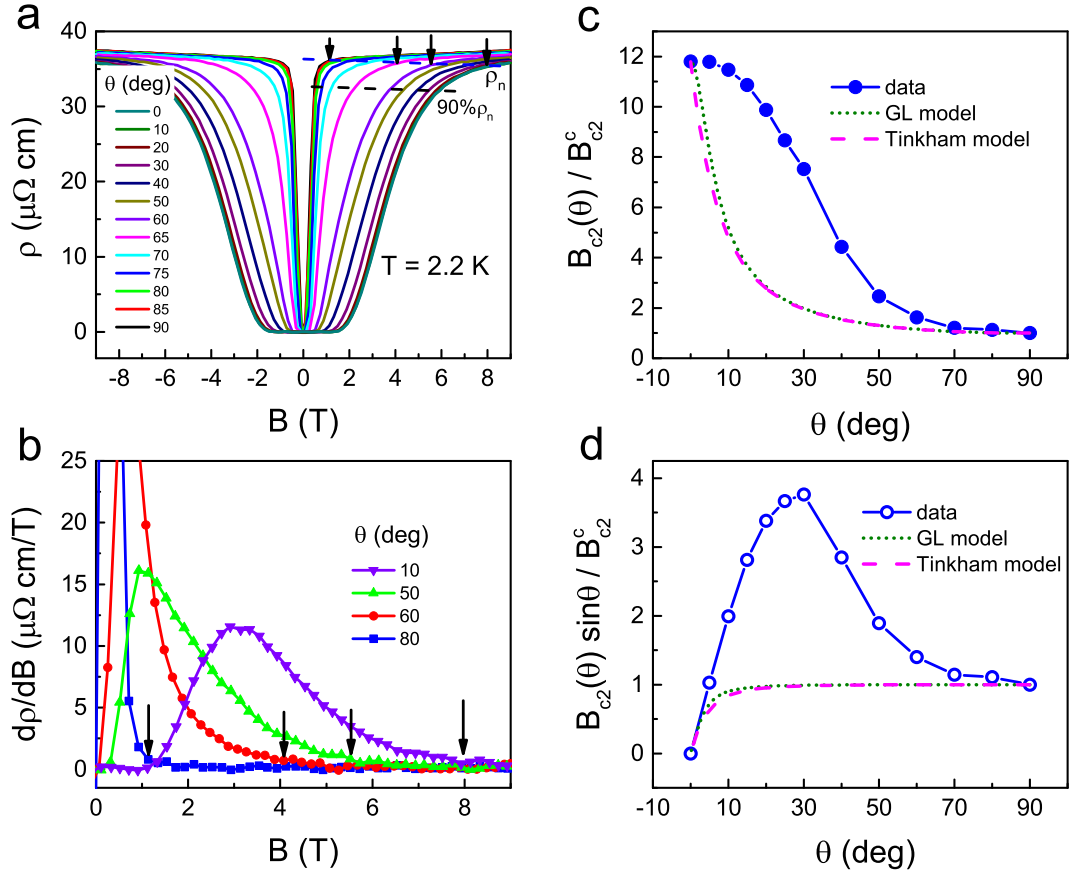}
\caption {The measurements of the upper critical field along
different directions $B_{c2}(\theta)$ for sample \#1. (a) Field
dependence of resistivity with the external field rotating from
$B\parallel ab$ ($\theta = 0^o$) to $B\parallel c$ ($\theta = 90^o$)
at a fixed temperature $T = 2.2$ K. (b) Differential of the $\rho-B$
curves in (a), based on which the onset superconducting transition
points are determined indicated by the black arrows. (c) Angular
dependence of the upper critical field $B_{c2}(\theta)$ normalized
by $B_{c2}^c$. (d) Field dependence of perpendicular component of
the upper critical field $B_{c2}(\theta) sin\theta$ normalized by
$B_{c2}^c$. In (c) and (d), the theoretical curves based on the GL
model and Tinkham model are shown in comparison with the
experimental data.} \label{fig2}
\end{figure*}

\begin{figure*}
\includegraphics[width=0.95\textwidth,bb=3 4 1438 574]{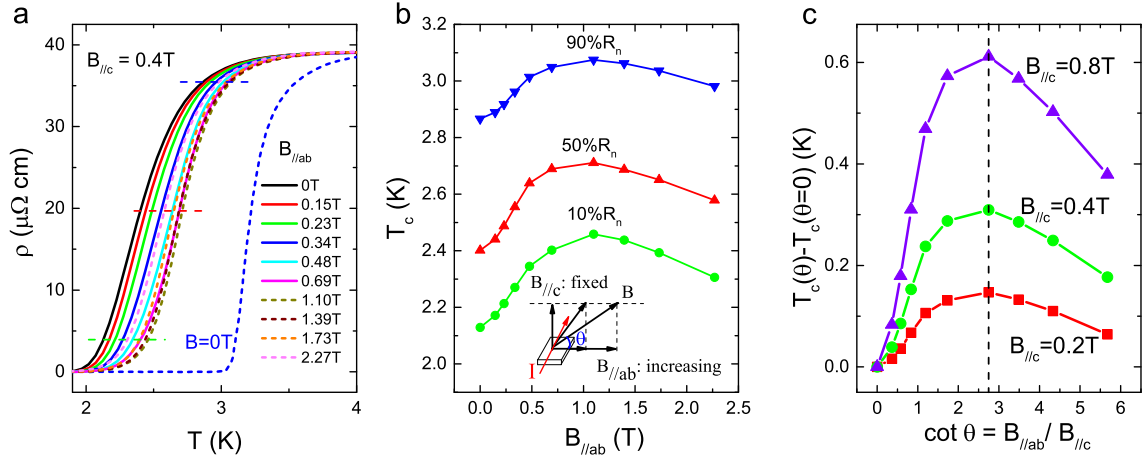}
\caption {Electrical resistivity data and the critical transition
temperature in the inclined field with $B_{\parallel c}$ fixed and
$B_{\parallel ab}$ increasing for sample \#2. (a) $\rho-T$ curves with
$B_{\parallel c}$ = 0.4 T and $B_{\parallel ab}$ increasing. (b)
$B_{\parallel ab}$ dependence of $T_c$ determined from the data in
(a) by three criterions 10\%$\rho_n$, 50\%$\rho_n$ and 90\%$\rho_n$. The inset shows a sketch map of the field configuration for the
measurements. $B_{\parallel ab}$ is applied perpendicular to the current. (c) $B_{\parallel ab}$ (normalized by $B_{\parallel c}$) dependence of the $T_c$ enhancement by the
criterion 50\%$\rho_n$ under three fixed $B_{\parallel c}$ = 0.2 T, 0.4
T and 0.8 T. } \label{fig3}
\end{figure*}

\begin{figure*}
\includegraphics[width=0.95\textwidth,bb=3 4 1078 591]{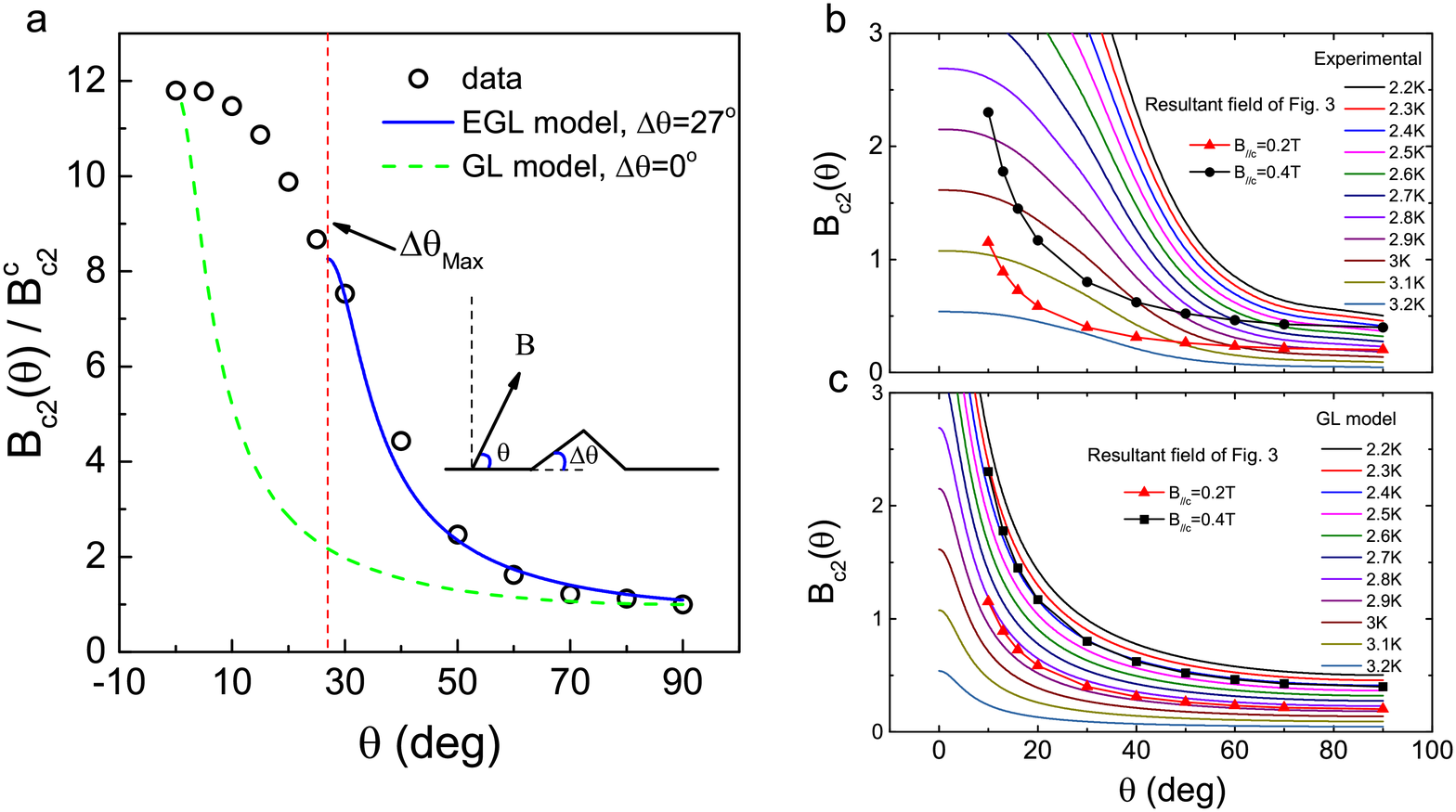}
\caption {(a) The comparison between the angle dependence of upper critical field and the extended GL (EGL) model considering the influence of the wrinkles.
(b) The comparison between the angle dependence of experimental upper critical field and the practical field applied in the experiments shown in Fig. 3.
(c) The comparison between the GL model and the practical field applied in the experiments shown in Fig. 3. } \label{fig3}
\end{figure*}

\begin{table}
\centering \caption{Summary of the upper critical fields
($B_{c2}^{ab}$ and $B_{c2}^{c}$) and the paramagnetic limiting field
($B_P$) of the sample \#1.}
\medskip
\begin{tabular}{cccccc}
\hline $T_c$ & $dB_{c2}^{ab}(T)/dT|_{T_c}$ & $dB_{c2}^{c}(T)/dT|_{T_c}$ & $B_{c2}^{ab}$ & $B_{c2}^{c}$  & $B_{P}$ \\
\hline
3.17 K & -4.34 T/K & -0.39 T/K & 9.5 T & 0.86 T & 5.8 T \\
\hline
\end{tabular}
\end{table}

\end{document}